\documentclass[%
 reprint,
 amsmath,amssymb,
 aps,
]{revtex4}

\pdfoutput=1
\usepackage{graphicx}
\usepackage{dcolumn}
\usepackage{bm}
\usepackage{tensor} 

\begin{document}

\title{Skewness of matter distribution in clustering dark energy cosmologies}

\author{Hermano Velten}%
 \email{hermano.velten@ufop.edu.br}
\affiliation{Departamento de F\'isica, Universidade Federal de Ouro Preto (UFOP), Campus Morro do Cruzeiro, 35400-000, Ouro Preto-MG Brazil}

\author{Raquel Emy Fazolo}%
 \email{raquel.fazolo@gmail.com}
\affiliation{ PPGCosmo, Universidade Federal do Esp\'irito Santo, 29075-910, Vit\'oria, ES, Brazil}

\date{\today}
             
\begin{abstract}
We calculate the skewness (the third moment $S_3$) of matter distribution in dynamical dark energy cosmologies. We pay particular attention to the impact of dark energy perturbations on this quantity. There is indeed a clear signature of dark energy perturbations on this quantity. By properly allowing dark energy perturbations we show that their impact on $S_3$ is strong enough (a factor $\sim 3$ greater) to easily discriminate between clustering and non-clustering dark energy cosmologies. This indicates that high order statistics of the cosmic density field are useful to the study of dark energy models and are potentially able to rule out clustering dark energy cosmologies.\\
{\bf Keywords}: Cosmology, dark energy, large scale structure \\
{\bf Pacs}: 98.80.-k, 95.36.+x, 04.25.Nx
\end{abstract}

\maketitle

\section{\label{sec:Intro}Introduction}

Dark energy is a key element of modern cosmology since it drives the late time accelerated background expansion of the universe. Although dark energy properties are usually studied via background related cosmological observables it is also possible to understand its nature since it acts indirectly on the large scale structure (LSS) of matter distribution by smoothing linear matter overdensities on scales beyond galaxy clusters. If dark energy appears in nature in the form of a cosmological constant $\Lambda$ it represents a smooth field and obviously does not cluster. On the other hand a dynamical dark energy field can eventually agglomerated depending on its rest frame speed of sound $c^{2}_{s\,DE} =\partial p_{DE} / \partial \rho_{DE}$, where $p_{DE}$ ($\rho_{DE}$) stands for the DE pressure (energy density). Hence, in dynamical dark energy cosmologies the quantity $c^{2}_{s\,DE}$ is a new degree of freedom at perturbative level and determines the clustering features of matter distribution. Due to the peculiarities of cosmological perturbation theory the first order DE perturbations in the case it is described by a scalar field (with intrinsic speed of sound $c^{2}_{s\,DE}=1$) oscillate and do not cluster. Therefore, due to the widely use of a quintessence scalar field representing dark energy \cite{Ratra:1987rm} one often neglects any manifestation of dark energy perturbations as a relevant source for LSS.     

Assuming the the perspective that dark energy perturbations can play a relevant role at first order the analysis of dark energy perturbations on LSS one can yield to reconstruction of the actual DE nature. In some approaches the DE speed of sound can evade the canonical value $c^{2}_{s\,DE}=1$ (see examples in Refs. \cite{Padmanabhan:2002cp,Bagla:2002yn,Garriga:1999vw,ArmendarizPicon:2000ah,Lim:2010yk,Creminelli:2008wc}) and therefore leaving imprints on structure formation  \cite{Mota:2004pa,Manera:2005ct,Abramo:2007iu,Abramo:2008ip,Pace:2014taa,Creminelli:2009mu,Basse:2010qp,Heneka:2017ffk,Sefusatti:2011cm,Fasiello:2016qpn,Fasiello:2016yvr,Batista:2017lwf,Endo:2018xhx}.

In this work we go beyond linear order in scalar perturbations by focusing on the skewness of matter distribution when DE perturbations are allowed. Skewness represents the third moment of matter fluctuations and provides a measurement on how the clustering process create an asymmetry between under/overdense regions.
Therefore since DE tends to increase the ``empty'' space between galaxies one expects that DE (and its perturbations) can play role in this process by modifying the expected balance between under/overdense regions with respect to non-clustering DE cosmologies. Indeed, the skewness for the case of a homogeneous dark energy component has been already studied long ago \cite{Peebles1980} by assuming that LSS forms by gravitational instabilities from initial Gaussian fluctuations. It is widely know that the skewness value has a weak dependence on the matter density parameters in the $\Lambda$CDM cosmology \cite{Bernardeau:1993qu,Bernardeau:1994vz}. Modified gravity theories \cite{Tatekawa:2008bw}, Chaplygin gas cosmologies \cite{Reis:2004hm} and even the equivalence principle \cite{Amendola:2004wa} have also been tested using the skewness. Non-clustering quintessence models have been constrained via the impact on the skewness \cite{Benabed:2001dm}. But to the best of our knowledge dark energy perturbations emerging from a dynamical field have not yet been considered into this calculation. This is the gap this work is aiming to fill. 

In the next section we present both the background and perturbative dynamics of our approach. We split the evolution of matter fluctuations into first and second order in perturbations. From this analysis we show in detail how the skewness is computed. In section III we present our results. We calculate the skewness for different dynamical DE cosmological models and a proper comparison of results is presented. We conclude in the final section IV.

\section{\label{sec:Dynamics}Background and Perturbed Dynamics}

We focus on late time homogeneous and isotropic background expansions in which a flat Friedmann-Lemaitre-Robertson-Walker (FLRW) metric gives rise to the expansion rate 
\begin{equation}
\frac{H^{2}(a)}{H^{2}_0}=\frac{\Omega_{\rm m0}}{a^{3}}+(1-\Omega_{\rm m0})\,\, e^{-3\int da\frac{1+w_{\rm DE}}{a}}.
\label{Ha}
\end{equation}
In the above equation $a$ is the scale factor with the expansion rate given by $H=\dot{a}/a$ where the symbol ``\,$\,^{.}\,$\,'' means derivative with respect to the cosmic time. The preferred cosmological parameters according to the recent Planck 2018 release are $\Omega_{\rm m0}= 8\pi G \rho_{m0}/3H^2_0=0.315\pm 0.007$ for the total matter fractionary parameter and $w_{DE}=-1.03\pm 0.03$ for the DE equation of state \cite{Aghanim:2018eyx}. The today's Hubble constant $H_0$ is indeed a free parameter of the background expansion but, as shown below, perturbations are not sensitive to this quantity. Then it is useless here to quote a value for $H_0$. In practise, for the dynamics of the late time Universe the radiation effects can be neglected since they sum up to $\Omega_{r0} \sim 10^{-5}$. Radiation has indeed a relevant role in structure formation at early times but does not impact the skewness distribution (the focus of this work) generated via gravitational instability at late times. We will also concentrate on alternative dynamical dark energy models which can be studied via a specific choice of the dark energy equation of state parameter $w_{DE}=p_{DE}/\rho_{DE}$. It can be either a constant parameter (being equivalent to a cosmological constant $\Lambda$ if $w_{DE}=-1$) or to assume a time dependence. The models used in this work will be described in the next section.

In order to investigate cosmological perturbations we adopt Newtonian-like equations for gravitating fluids as done in Refs. \cite{Lima:1996at,Reis:2003fs,Reis:2004hm,Abramo:2007iu,Abramo:2008ip,Fabris:2008hy,Velten:2011ab,Batista:2013oca,Pace:2014taa}. This is a valid approximation since we are focusing on the mass density distribution on scale well within the horizon. In this approach the mass density conservation is expressed by the continuity equation as
\begin{equation}
\Big{(}\frac{\partial\rho}{\partial t}\Big{)}_{r}+\vec{\nabla}_{r}\cdot(\rho\vec{u})+P\vec{\nabla}_{r}\cdot\vec{u}=0.
\label{cont}
\end{equation}
Actually the above equation corresponds to mass conservation in addition to the pressure $P$ effects on this balance. 

The Euler equation provides the momentum conservation for a fluid sourced by the gravitational potential $\phi$, 
\begin{equation}
    \Big{(}\frac{\partial\vec{u}}{\partial t}\Big{)}_{r}+(\vec{u}\cdot\vec{\nabla}_{r})\vec{u}=-\vec{\nabla}_{r}\phi-(\rho+P)^{-1}\vec{\nabla}_{r}P.
\label{ueq}
\end{equation}

Finally, the Poisson equation identifies the source of gravitational field with the active gravitational mass density $(\rho +3P)$ such that
\begin{equation}
    \nabla_{r}^{2}\phi=4\pi G(\rho+3P).
\label{poisson}
\end{equation}

Eqs. (\ref{cont})-(\ref{poisson}) are called in the literature neo-Newtonian/pseudo-Newtonian equations. When sourced by a velocity field corresponding to the Hubble expansion law $\vec{u}=H \vec{r}$ they are consistent with the background dynamics (Friedmann equations) obtained in General Relativity. This gives rise to the so called neo-Newtonian cosmology \cite{neoNewton1,neoNewton2,neoNewton3}. Hence, although our aim in using (\ref{cont})-(\ref{poisson}) will be related to the perturbations this set is self-consistent with the use of the background expansion rate given by Eq. (\ref{Ha}).

The inclusion of the pressure term in these equations represents a new aspect in comparison to standard hydrodynamics. The term $P \vec{\nabla}_{r} \cdot \vec{u}$ in Eq. (\ref{cont}) has been introduced in Ref. \cite{Lima:1996at} in order to match linear cosmological scalar matter perturbations derived from the set (\ref{cont})-(\ref{poisson}) with general relativity predictions. See also \cite{Abramo:2007iu,Reis:2003fs}

Scalar cosmological perturbations can be studied with Eqs.(\ref{cont})-(\ref{poisson}) via the introduction of small fluctuations around the background quantities. A perturbative expansion for the mass overdensities $\delta=\delta \rho / \rho$ and the peculiar velocity $\vec{v}$ is obtained by linearizing the system  (\ref{cont})-(\ref{poisson}) with 
\begin{eqnarray}
\rho=\rho_{0}+\delta\rho, \,\, P=P_{0}+\delta P, \,\, \phi=\phi_{0}+\varphi, \,\,\vec{u}=\vec{u}_{0}+\vec{v}.
\label{pert1}
\end{eqnarray}

The use of a comoving coordinate system ($\vec{x}=a^{-1}\vec{r}$) allows to properly follow the evolution of perturbed quantities. Given the change of coordinates and the corresponding operators $\vec{\nabla}_{x}=a\vec{\nabla}_{r}$ and $(\partial_t)_{x}=(\partial_r)_{r}+H(\vec{x}\cdot\vec{\nabla}_{x})$, with introduction of (\ref{pert1}) into Eqs.(\ref{cont})-(\ref{poisson}) and after collecting perturbed terms up the second order the perturbed version of Eqs.(\ref{cont})-(\ref{poisson}) reads, respectively,
\begin{equation}
    \dot{\delta}+3H(c_{eff}^{2}-w)\delta+\frac{(1+c_{eff}^{2})}{a}(\vec{\nabla}\cdot\vec{v})\delta+\frac{1+w}{a}\vec{\nabla}\cdot\vec{v}=0,
\label{contpert2}\end{equation}

\begin{equation}
    \dot{\vec{v}}+H\vec{v}+\frac{1}{a}\vec{\nabla}\varphi+\frac{\vec{\nabla} c_{eff}^{2}\delta}{a\left[1+w+(1+c^2_{eff})\delta\right]}+\frac{(\vec{v}\cdot\vec{\nabla})}{a}\vec{v}=0,
\label{Eulerpert2}
\end{equation}

\begin{equation}
    \nabla^{2}\varphi=4\pi Ga^{2}\rho_{0}\delta(1+3c_{eff}^{2}).
\label{poissonpert2}\end{equation}

In the above equations we have defined $w=P_0/\rho_0$ and the effective speed of sound $c_{eff}^{2}=\delta P / \delta\rho$. It is worth noting that both first order and second order perturbations are present in Eqs. (\ref{contpert2})-(\ref{poissonpert2}). The aim now is to obtain a single differential equation for the evolution of density fluctuations. Hence, by taking the time derivative of Eq. (\ref{contpert2}), the divergence of Eq.(\ref{Eulerpert2}) and combining them with (\ref{poissonpert2}) one obtains the following equation for the density contrast

\begin{eqnarray}
&&a^{2}\delta^{\prime \prime}+a\delta^{\prime}\left[\frac{aH^{\prime}}{H}+3-3w+3c_{eff}^{2}-\frac{w^{\prime}a}{1+w+\delta(1+c_{eff}^{2})}\right]
\nonumber \\
&&-\delta^{\prime 2}a^{2}\left[\frac{4/3+c_{eff}^{2}}{1+w+\delta(1+c_{eff}^{2})}\right]+\delta\delta^{\prime}a\left[\frac{(w-c_{eff}^{2})(5+3c_{eff}^{2})-c_{eff}^{2\prime}a}{1+w+\delta(1+c_{eff}^{2})}\right]\nonumber \\
&&+\delta^{2}\left[\frac{3(w-c_{eff}^{2})c^{2\prime}_{eff}a-3(w-c_{eff}^{2})^{2}}{1+w+\delta(1+c_{eff}^{2})}-\frac{3}{2}(1+3c_{eff}^{2})(1+c_{eff}^{2})\right] \\ \nonumber
&&+\frac{3}{2}\delta\left[(3w^{2}-2w-1)-2c_{eff}^{2}(1+3w)-2(w^{\prime}a-c_{eff}^{2 \prime}a)+\frac{2(w-c_{eff}^{2})w^{\prime}a}{1+w+\delta(1+c_{eff}^{2})}\right]= \\ \nonumber
&&\frac{1+w+\delta(1+c_{eff}^{2})}{a^{2}H^{2}}\nabla_{x}\left(\frac{\nabla_{x}(c_{eff}^{2}\delta)}{1+w+\delta(1+c_{eff}^{2})}\right).
\label{fulldelta}
\end{eqnarray}

In the above equation the symbol ($^{\prime}$) means derivative with respect to the scale factor $a$. Eq. (\ref{fulldelta}) contains up to second order terms in the matter density contrast for a single fluid fluid with equation of state parameter $w$ and effective speed of sound $c^2_{eff}$. A similar Eq. has been obtained in Ref. \cite{Reis:2004hm} where the skewness of Chaplygin gas models have been investigated. 

The calculation of the skewness follows the principles presented in Refs. \cite{Goroff:1986ep,Bernardeau:1993qu}. One has to calculate the leading order of the third moment of matter density distribution $S_3 = \left\langle \delta^3 \right\rangle / \left\langle \delta^{2}\right\rangle^2$ where $\left\langle ...\right\rangle $ means an ensemble average. This calculation is done by  expanding the density contrast $\delta$ with respect to a small perturbation $\delta_0$ which corresponds to a initial density fluctuation. Then, in order study the evolution of the density contrast at first and second perturbative level we apply the decomposition
\begin{equation}
    \delta = \sum^{\infty}_{i=1}\delta_{i}=\sum^{\infty}_{i=1}\frac{D_{i}(\eta)}{i!}\delta^{i}_{0}.
\label{expansiondelta}\end{equation}
This allows us to split Eq. (\ref{fulldelta}) in homogeneous equation for the first order perturbation $D_{1}$:

\begin{eqnarray}
&&a^{2}D_{1}^{\prime\prime}+aD_{1}^{\prime}\left(\frac{aH^{\prime}}{H}+3-3w+3c_{a}^{2}-\frac{aw^{\prime}}{1+w}\right)\nonumber \\
&&-\frac{3}{2}D_{1}\left(1+2w-3w^{2}+2c_{a}^{2}(1+3w)+2(aw^{\prime}-c_{a}^{2})-\frac{2aw^{\prime}(w-c_{a}^{2})}{1+w}\right)=0.
\label{D1}
\end{eqnarray}

Since there are only adiabatic modes the effective speed sound $c^2_{eff}$ equals its adiabatic component $c^2_{a}=\partial P / \partial \rho$. The above equation can be solved numerically since an EoS is provided. Then, the solution for $D_{1}$ is used as a source term in the equation for the second order perturbation $D_{2}$:

\begin{eqnarray}
&&a^{2}D_{2}^{\prime\prime}+aD_{2}^{\prime}\left(\frac{aH^{\prime}}{H}+3-3w+3c_{a}^{2}-\frac{aw^{\prime}}{1+w}\right)-\frac{3}{2}D_{2}\left(1+2w-3w^{2}+2(w^{\prime}a-c_{a}^{2})-\frac{2aw^{\prime}(w-c_{a}^{2})}{1+w}\right)= \nonumber \\
&&a^{2}D_{1}^{\prime 2}\left(\frac{8/3+c_{a}^{2}}{1+w}\right)-aD^{\prime}_{1}D_{1}\left(\frac{2(w-c_{a}^{2})(5+3c_{a}^{2})-a^{2}c_{a}^{2}}{1+w}+\frac{2aw^{\prime}}{(1+w)^{2}}+\frac{(8/3)c_{a}^{2}}{(1+w)^{2}}\right)\nonumber \\
&&+3D_{1}^{2}\left((1+3c_{a}^{2})(1+c_{a}^{2})+\frac{2(w-c_{a}^{2})^{2}-2(w-c_{a}^{2})a^{2}c_a^{ 2\prime}}{1+w}+\frac{2a(w-c_{a}^{2})w^{\prime}}{(1+w^{2})}\right).
\label{D2}
\end{eqnarray}
Both equations (\ref{D1}) and (\ref{D2}) have the correct limit for a pressureless fluid as used in Refs. \cite{Bernardeau:1994vz,Benabed:2001dm,Gaztanaga:1997ah}.

With the solution of Eqs. (\ref{D1}) and (\ref{D2}) one obtains the relevant quantity for the skewness distribution
\begin{equation}
S_3 = \frac{3D_2}{D_1^2}.
\label{S3}
\end{equation}

It is worth noting that the system (\ref{D1}) and (\ref{D2}) has been obtained for a single gravitating fluid (with pressure $P$ and speed of sound $c^2_a$) described by (\ref{cont})-(\ref{poisson}). At this point the quantities $H$, $w$ and $c^2_a$ have to be specified in order to study a specific model. 

\section{Results}

In the last section we provided the equations for the evolution of density fluctuations $\delta$ up to its second order in the expansion (\ref{expansiondelta}). This allows us now to compute the quantity $s_3$ as in (\ref{S3}) for a giving cosmology. The strategy is to solve numerically equation (\ref{D1}) for $D_1$. In this step the initial conditions $\delta(a_0)=10^{-3}$ and $\delta^{\prime}(a_0)=0$ are used where $a_0=10^{-3}$ is the scale factor around the decoupling. The solution found for $D_1$ is then later used as a source term in Eq. (\ref{D2}). We finally solve numerically Eq. (\ref{D2}) and calculate today's value for $S_3$. In order to apply Eqs. (\ref{D1}) and (\ref{D2}) - which have been obtained for a one component fluid with density $\rho$ and pressure $P$ - one has to properly interpret the quantity $\delta$ and then specify the background expansion rate $H$. For the latter we focus on expansion rates as in Eq. (\ref{Ha}) and different cosmological models are given by particular choices of the parameters $\Omega_{m0}$ and $w_{DE}$. 

The Einstein de-Sitter universe is modelled as composed by a single pressureless ($P=0$) matter fluid throughout the entire cosmic expansion i.e., $\Omega_{m0}^{EdS}=1$ $\forall a$. It surely fails to account for dark energy effects but it is indeed a realistic description for the background expansion in the redshift range $z_{eq} < z < z_{DE}$ i.e., the period between the radiation/matter equality ($z_{eq} \sim 3300$) until dark energy effects take place at redshifts $z_{DE}\lesssim 1$. Within this period structure formation take place. Using such parameter values into the systems (\ref{D1}) and (\ref{D2}) we correctly find the value $S_3^{EdS}=34/7$ (see Ref. \cite{Bernardeau:1993qu}) when evolving this set of equation until today. In a realistic universe however one has $\Omega_{m0}\sim 0.3$. Also, the dark energy sector has to be modelled both via specifying its equations of state parameter $w_{DE}$ and providing a proper description of its density perturbation $\delta \rho _{DE}$ behavior.    

\begin{figure}
\includegraphics[width=0.475\textwidth]{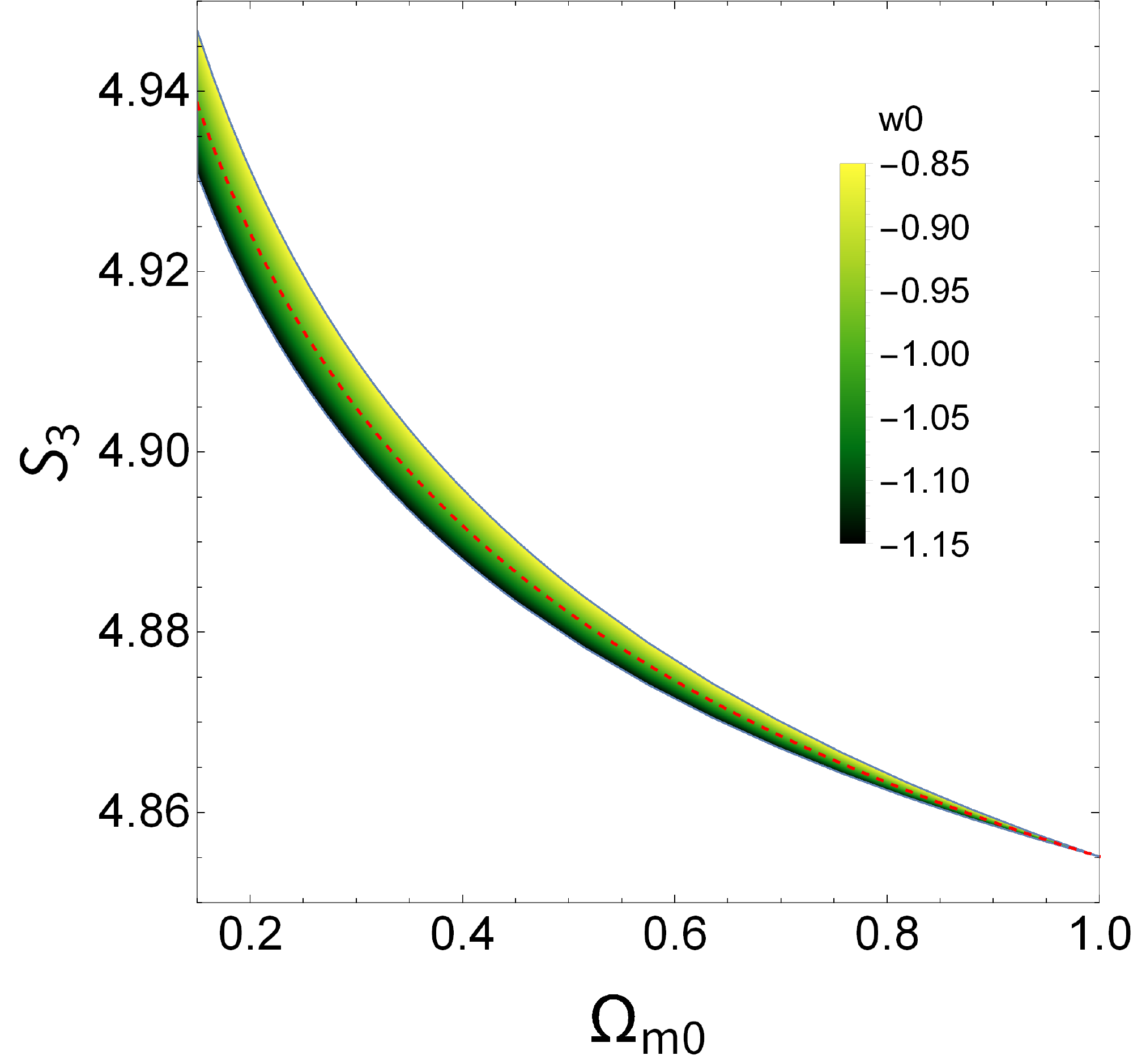}
\includegraphics[width=0.465\textwidth]{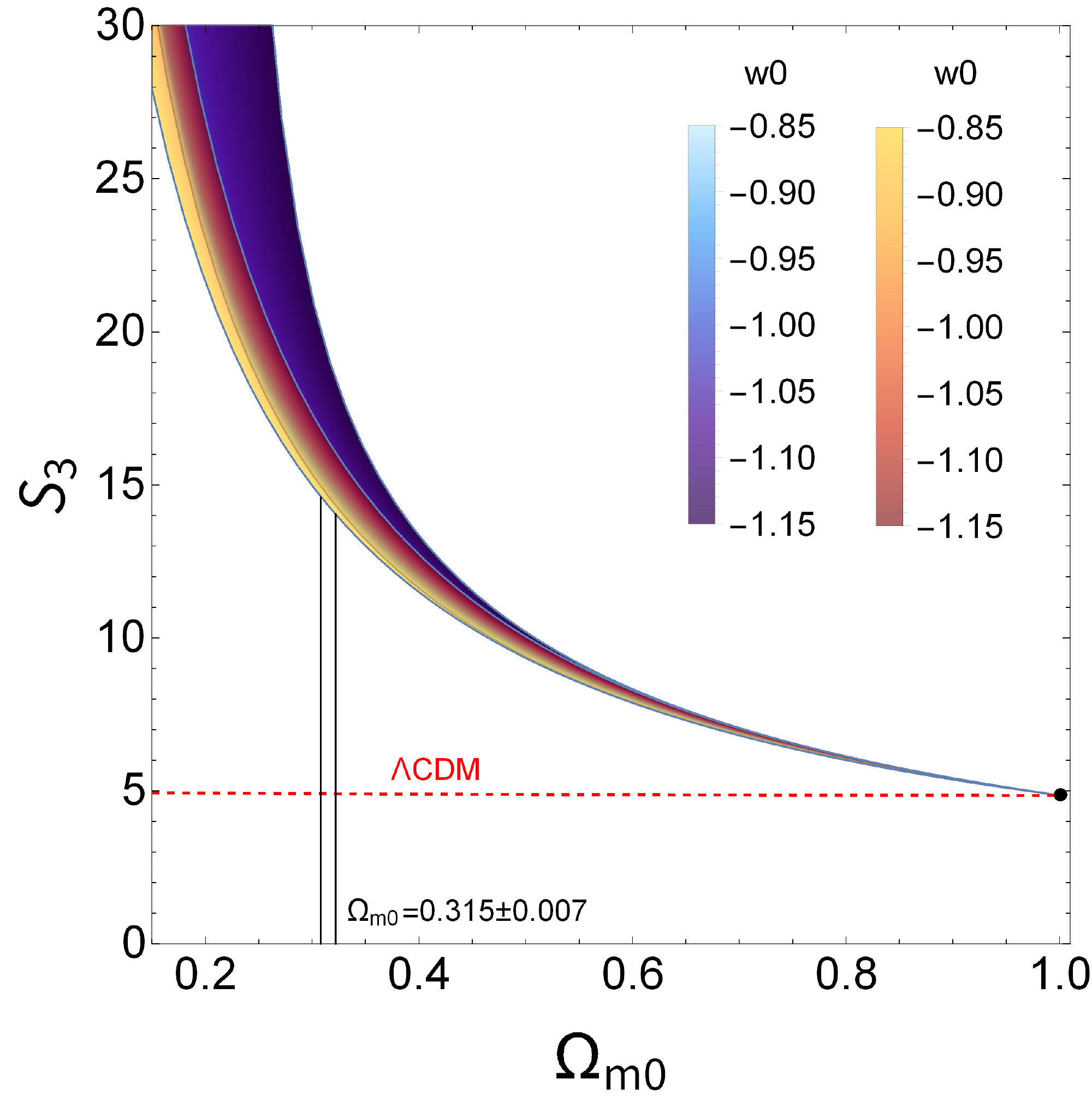}
\caption{Skewness as a function of the today's fractionary matter density $\Omega_{m0}$. Left panel: Dark energy perturbations are not considered. The LCDM curve is denoted by the red dashed line. The skewness assumes $S_3\approx 4.91$ for a $\Omega_{m0}=0.315$ universe. We add the information on the impact of the dark energy equation of state. The stripe considers $w$CDM cosmologies without any dark energy perturbations with $w_0$ values given according to the color scheme shown in the inset of this panel. $w_0$ values $< -1 (>-1)$ decreases (increases) $S_3$ but the impact is quite weak. Right panel: Dark energy perturbations are considered here. For $\Omega_{m0}=0.315\pm 0.07$ values as given by the Planck best fit (which can be visualized by the vertical lines) the skewness assumes values $\sim 15$. The blue stripe also considers dark energy perturbations but its speed of sound is set to zero $c^{2}_{a}=0$. The black dot at $\Omega_{m0}=1$ represents the Einstein the Sitter model ($\Omega_{m0}=1$) with its standard value $34/7\sim 4.857$. The $\Lambda$CDM universe corresponds to the dotted red line.}
\label{FigSkewness}
\end{figure}

We show the core results of this work in Fig. \ref{FigSkewness}. The panels show today's $S_3$ as a function of the fractionary matter density parameter $\Omega_{m0}$. The left panel of this figure focus on the standard cosmological scenarios and it amplifies the region around the $\Lambda$CDM model (the red dashed line in both panels). In order to obtain the stripe in this panel we have assumed a single pressureless $w\rightarrow w_{m}\equiv0$ matter fluid i.e. we identify $\delta \rightarrow \delta_m$ in Eqs. (\ref{D1}) and (\ref{D2}). Consequently the speed of sound of the matter fluid vanishes $c^2_a=0$. Dark energy effects enter here only via their contribution to the background expansion as given by (\ref{Ha}). This means that there are no dark energy perturbations in this scenario.  For the flat $\Lambda$CDM model (the red dashed line) there is only a slightly variation from $S_3=34/7=4.857$ (as expected in a Einstein-de Sitter model when $\Omega_{m0}=1$) to values $S_{3}\sim 4.91$ considering $\Omega_{m0} \sim 0.3$. 
For different dark energy models in which $w_{DE} \neq -1$ the impact $w_{DE}$ is seen by the color gradient in the stripe following the color scheme in the inset of this panel. Phantom regimes ($w_{DE}<-1$) slightly decreases the $S_3$ value (dark region in this stripe) while for $w_{DE}>-1$ (the yellow region of the stripe) increases $S_3$. This is in agreement with previous results obtained in e.g., Ref. \cite{Benabed:2001dm}. We have also verified that for a time dependent dark energy equation of state like the CPL model 
$w_{DE}=w_{0}+w_{1}(1-a)$ the $S_3$ values remain similar to the expected $S_3 <5$ upper limit on the skewness parameter for a large range of $w_1$ parameter values. 

We assess now the impact of DE perturbations $\delta_{de}$ on the $S_3$. Our strategy is to calculate (\ref{D1}) and (\ref{D2}) by interpreting $\rho$ as the total energy density of the system (i.e., the sum of matter with dark energy). Hence, at perturbative level the total density fluctuation will be affected by both standard matter fluctuations $\delta_m$ as well as by dark energy perturbations $\delta_{de}$. Indeed, the effective cosmological clustering due to the gravitational interaction measured by cosmic surveys make no distinction between $\delta_m$ and $\delta_{DE}$ since both will source the gravitational potential $\phi$. The actual quantity measured is the total clustering which in the standard cosmology (where $\delta_{DE}$ is absent) is associated only to baryonic fluctuations ($\delta_B$) within a dark matter potential well which is related to the dark matter overdensity $ \delta _{DM}$. Although these components can be quantitatively distinguished via the introduction of a bias factor $b$ such that $\delta_B = b \delta _{DM}$ they both contribute to the total matter perturbation $\delta_{m}$. Therefore, this justifies the use of the total fluid fluctuation as an effective way to include dark energy clustering. Similar approaches for studying the impact of dark energy perturbations to the total fluid perturbative dynamics has been implemented for example in Refs. \cite{Dent:2008ek,Velten:2017mtr,Velten:2018bws,Sefusatti:2011cm}. The difference now is that while in standard cosmology baryons and dark matter are considered pressureless fluids (and therefore the total matter fluid is pressureless) the situation is different when dark energy is added to the total density perturbation. Indeed, the effective (with respect to the total fluid) equation of state parameter and the speed of sound will differ from the standard pressureless case. This implies in defining an effective total fluid EoS parameter 
\begin{equation}
    w(a)=\sum \Omega_{i} (a)w_{i}(a)=\frac{\Omega_{de0}H_0^2}{H^2(a)}e^{-3\int da\frac{1+w_{de}}{a}}w_{de}(a).
\end{equation}
The above relation is obtained assuming a total pressure $p=p_m+p_{de}$ but considering  $p_{m}=0$. The total density reads $\rho=\rho_{m}+\rho_{de}$. The adiabatic speed of sound of the total fluid is then written as
\begin{equation}
c_{\rm a}^{2}=\frac{\dot{p}}{\dot{\rho}}=\frac{w}{1+w}\Big{[}(1+w_{\rm DE})-\frac{a}{3}\frac{w^{\prime}_{\rm DE}}{w_{\rm DE}}\Big{]}.
\end{equation}

Since we calculate numerically now Eqs. (\ref{D1})-(\ref{D2}) with $c^{2}_s \neq 0$ there is a scale dependence on our results appearing from the $\Delta_x$ operators (\ref{D2}) i.e., $\nabla^2_x \rightarrow - k^2 /a^2$. We set $k=0.1$ h/Mpc since it corresponds to scale that are still within the linear regime at $z=0$ and most attempts to measure higher order moments of the large scale density field targets on similar scales \cite{Szapudi:2001mh,Mao:2014caa}.

 The right panel of Fig.1 shows the $S_3$ value as a function of $\Omega_{m0}$ when dark energy perturbations are taken into account. We find that dark energy perturbations yields to a substantial increase in $S_3$. The dark stripe (with color scheme violet/light blue) of this figure considers the presence of dark energy perturbations for cosmological models given by Eq. \ref{Ha} when the dark energy equation of state parameters assumed values in the range $-1.15 < w_0 < -0.85$. The colour gradient shown in the insets of this panel provides information on the impact of changing $w_0$. It is worth noting that the $S_3$ values for $\Omega_{m0} \sim 0.315$ (the PLANCK 2018 preferred value) are around $\sim 17$ which is about a factor 3 greater than the standard EdS value or the standard $\Lambda$CDM cosmology (when dark energy perturbations are not considered) as shown in the left panel of this figure. This panel also shows a second stripe for which we have arbitrarily set $c^2_a=0$ as done in Refs. \cite{Creminelli:2009mu,Batista:2017lwf}. In this case the color scheme used to display the impact of $w_0$ ranges from red ($w_0=-1.15$) to yellow ($w_0=-0.85$). There is a tiny superposition between both colored stripes. Results of right panel are also weakly dependent on variations of the parameter $w_1$. Then, the skewness poorly constrains the time dependence of $w_{DE}$.

\section{Conclusions}

The actual physical process behind the accelerated cosmic expansion is one of the main questions to be faced in the next decades. A cosmological constant (and therefore a non-clustering component) is still favoured by current cosmological data \cite{Aghanim:2018eyx} but faces serious theoretical problems \cite{Weinberg:1988cp}. If dark energy is represented by a dynamical field there is a obvious time dependent behavior at the background level but one may also wonder what are the clustering features of this field.

We have considered the impact of dark energy perturbations by calculating the scalar density perturbations up to second order and then computing the skewness -  Eq. (\ref{S3}) - of matter distribution. For non-clustering dark energy cosmologies $S_3$ is weakly dependent on both the $\Omega_{m0}$ parameter and the equation of state parameter (with values as shown in the left panel of Fig.1). While the standard result $S_3=4.857$ is obtained for the Einstein-de Sitter model one finds $S_3 \sim 4.91$ for the standard $\Lambda$CDM cosmology. 

By including dark energy perturbations on the other hand we found a relevant impact on the skewness which assumes values in the range  $15 \lesssim S_3 \lesssim 20$ depending on the dark energy equation of state parameter and whether or not the dark energy speed of sound is considered. By assuming that there are no extra source of non-gaussianity which could lead to large $S_3$ values it is clear from Fig. 1 the intense impact of dark energy perturbations on the skewness value. This should occurs due to the fact that in our treatment both dark matter and dark energy have been considered as a single fluid. Therefore, this means that dark energy effectively clusters the matter component. The case of a non-clustering dark energy field is trivially visualized in the left panel of Fig. 1. Our results show that an accurate measurement of the skewness can shed light on a fundamental question cosmology research related to whether or not dark energy clumps. The answer for this question should appear by joining efforts from both the theoretical and observational sides as well as via LSS numerical simulations as done recently in \cite{Dakin:2019vnj}. 

Although our results indicate a relevant  sensitivity of high order statistics of the cosmic density field when dark energy perturbations are present we remark that a more detailed study taken into account a separate baryonic component should be carried out. Actual measurements observed galaxies in cell counts and therefore a baryonic component has to be added to the the dynamics of Section 2. In particular, the relative nonadiabatic perturbations between the dark fields (dark matter and dark energy) and the baryonic distribution can provide further observable consequences as shown in \cite{Winfried}. This would provide a more realistic cosmological observable for the skewness component of matter distribution.

\noindent
\section*{Acknowledgments}
 HV thanks CNPq and PROPP-UFOP for partial financial support. RF thanks FAPES for financial support. We thank Luca Amendola for interesting remarks on the first version on this work.

\end{document}